# Antifragility Predicts the Robustness and Evolvability of Biological Networks through Multi-class Classification with a Convolutional Neural Network

**Hyobin Kim** [1,2], **Stalin Muñoz** [3], **Pamela Osuna** [4], **and Carlos Gershenson** [5,6,7,*]

[1] Biotech Research and Innovation Centre (BRIC), University of Copenhagen (UCPH), Copenhagen, Denmark
[2] Novo Nordisk Foundation Center for Stem Cell Biology, DanStem, Faculty of Health Sciences, University of Copenhagen, Copenhagen, Denmark
[3] Institute for Software Technology (IST), Graz University of Technology, 8010 Graz, Austria
[4] Faculté des Sciences et Ingénierie, Sorbonne Université, 4 place Jussieu 75005 Paris, France
[5] Centro de Ciencias de la Complejidad, Universidad Nacional Autónoma de México, 04510 CDMX, México
[6] Instituto de Investigaciones en Matemáticas Aplicadas y en Sistemas, Universidad Nacional Autónoma de México, 04510 CDMX, México
[7] ITMO University, St. Petersburg, 199034, Russian Federatio
* Correspondence: cgg@unam.mx

**Abstract:** Robustness and evolvability are essential properties to the evolution of biological networks. To determine if a biological network is robust and/or evolvable, it is required to compare its functions before and after mutations. However, this sometimes takes a high computational cost as the network size grows. Here we develop a predictive method to estimate the robustness and evolvability of biological networks without an explicit comparison of functions. We measure antifragility in Boolean network models of biological systems and use this as the predictor. Antifragility occurs when a system benefits from external perturbations. By means of the differences of antifragility between the original and mutated biological networks, we train a convolutional neural network (CNN) and test it to classify the properties of robustness and evolvability. We found that our CNN model successfully classified the properties. Thus, we conclude that our antifragility measure can be used as a predictor of the robustness and evolvability of biological networks.

## 1. Introduction

Robustness and evolvability are prevalent in the evolution of biological systems [1-6]. As studying the relationship between the two properties is necessary for understanding how biological systems can withstand mutations and simultaneously generate genetic variations, numerous studies on their relationship have been done [7-11]. Robustness allows the existing functions to be preserved in the presence of mutations or perturbations, while evolvability enables new functions to be expressed for adapting to new environments [12-14]. To determine if a biological system is robust and/or evolvable, the comparison of its functions before and after internal perturbations is needed. Pragmatically, in Boolean networks used as gene regulatory network models, the definition of robust and evolvable networks has been established via the comparison of dynamic attractors (*i.e.*, stable steady states) before and after internal perturbations [14] based on numerical and experimental evidence showing that the attractors represent cell types or cell functions [15-17].

The definition of robustness and evolvability has been applied to a number of studies adopting Boolean network models in artificial life and systems biology [18-20]. However, the calculations for finding all attractors of the networks are computationally expensive. This is because as the network size $N$ grows, there is a combinatorial explosion of the state space (*i.e.*, the set of all possible states, whose size is $2^N$). As an alternative to this exhaustive approach, some studies where large networks are explored only sample the state space. This implies that only the attractors with the largest basins (*i.e.*, the sets of states leading to attractors) are found [14, 19, 21]. However, in the case of extremely long attractor lengths, many number of attractors, and/or evenly distributed basins of attraction, this strategy will give biased results. Here we aim to develop a predictor of robustness and evolvability without the explicit comparison of the functions (*i.e.*, attractors), and thus applicable to large networks.

We use antifragility to estimate the robustness and evolvability of biological networks. Antifragility can be defined as the ability of a system to improve its functionality in the presence of external perturbations [22]. In the context of Boolean networks, it can be measured by our previous approach [23, 24], where antifragility is easily calculated by means of complexity computed from partial state transitions. With the differences of antifragility before and after internal perturbations, we train a CNN model and then test it to classify the properties of robustness and evolvability. We found that our model successfully classified both properties. Thus, we conclude that antifragility can be used as a significant predictor of robustness and evolvability.

Our predictor — antifragility — has many potential applications. It would be useful to systems and computational biologists studying the properties of large biological networks from a dynamical perspective. They could find out if the large networks have the potential of being either robust or evolvable, or both without investigating how the functions (attractors) of the networks are changed before and after mutations. Also, understanding antifragility would be helpful for uncovering the mechanism of how biological systems acquired robustness and evolvability. Besides, our antifragility measure could be used as a control parameter to build robust and/or evolvable engineered systems.

## 2. Materials and Methods

### 2.1. Boolean Networks and Biological Systems

Boolean networks were proposed by Stuart Kauffman as gene regulatory network models [25-27]. They have been extensively used in many areas including artificial life, robotics, and systems biology [28-34]. They consist of nodes and links, where nodes represent genes, and the links represent interactions between genes. Each node has a binary state 0 (OFF) or 1 (ON). 0 means being inhibited, and 1 indicates being activated or expressed. The future state of each node is determined by a lookup table. The states of nodes are updated synchronously in discrete time. Once the network topology and the lookup tables for each node are set, they remain fixed.

If the links are randomly arranged, and the node states are updated by Boolean functions randomly assigned to each node, the networks are called Random Boolean networks (RBNs). RBNs are also known as Kauffman's *NK* Boolean networks, where *N* is the number of nodes and *K* is the number of links per node (self-links can be included). Meanwhile, the links and the update rules of Boolean network models of biological systems are determined by experiments and literature that have identified actual relationships between genes (or proteins).

In Figure 1, an example RBN and its state transition diagram are presented. With a Boolean network with N nodes, its state transition diagram graphically represents $2^N$ states and transitions among them. In the state transition diagram, there are two important concepts: one is *attractors*, and the other is *basins of attraction*. Attractors are states in a fixed-point or a limit cycle, and basins of attraction are the rest of the configurations going toward attractors. For a better understanding, we provide formal definitions of a Boolean network, an attractor, and a basin of attraction as follows [35]:

- We denote a *Boolean network* as $G(V, F)$ composed of a set $V = \{v_1, v_2, ..., v_N\}$ of nodes and a list $F = (f_1, f_2, ..., f_N)$ of Boolean functions, in which $f_i(v_{i_1}, v_{i_2}, ..., v_{i_k})$ is the Boolean function assigned to node $v_i$ and has input nodes $v_{i_1}, v_{i_2}, ..., v_{i_k}$. $IN(v_i)$ signifies the set of input nodes $v_{i_1}, v_{i_2}, ..., v_{i_k}$ to $v_i$.

- The state of each node is 0 or 1 at discrete time $t$. We use $v_i(t)$ to indicate the state of node $v_i$ at time $t$. Then, the state of node $v_i$ at time $t+1$ is denoted by $v_i(t+1) = f_i(v_{i_1}(t), v_{i_2}(t), \ldots, v_{i_k}(t))$.
- If we designate $[v_1(t), v_2(t), \ldots, v_N(t)]$ as $\boldsymbol{v}(t)$, $v_i(t+1)$ is simply written by $v_i(t+1) = f_i(\boldsymbol{v}(t))$. $\boldsymbol{v}(t)$ means a gene expression profile at time $t$. For the whole Boolean network, it is written by $\boldsymbol{v}(t+1) = \boldsymbol{f}(\boldsymbol{v}(t))$.
- The set of edges $E$ is defined as $E = \{(v_{i_j}, v_i) | v_{i_j} \in IN(v_i)\}$. Then, $G(V, E)$ represents a directed graph which shows the network topology. An edge from $v_{i_j}$ to $v_i$ means that $v_{i_j}$ has an effect on the expression of $v_i$. $K$ is the in-degree of $v_i$.
- An initial $\boldsymbol{v}(0)$ eventually converges into a set of state configurations, which is defined as an *attractor*. When an attractor is composed of only one state configuration (*i.e.*, $\boldsymbol{v} = \boldsymbol{f}(\boldsymbol{v})$), it is a *fixed-point attractor*. When an attractor is composed of more than one state configuration, it is a *cyclic attractor* with period $l$ if it consists of $l$ state configurations (*i.e.*, $\boldsymbol{v}^1 = \boldsymbol{f}(\boldsymbol{v}^l) = \boldsymbol{f}(\boldsymbol{f}(\boldsymbol{v}^{l-1})) = \cdots = \boldsymbol{f}(\boldsymbol{f}(\cdots \boldsymbol{f}(\boldsymbol{v}^1)\cdots)))$. Here, $l$ is called *attractor length*. The set of state configurations which eventually reach the same attractor is defined as *basin of attraction*. For example, in Figure 1, 000, 010, and 111 are fixed-point attractors. {011, 110} is a cyclic attractor with $l = 2$. {100, 101} belong to the basin of the cyclic attractor {011, 110}.

In this study, we use different Boolean network models of 37 biological systems. The range of the network size is from 5 to 26. We find all attractors and basins of attraction by exhaustively tracking which states all state configurations finally converge to. We use Dubrova and Teslenko's algorithm to find attractors [75]. Therefore, to find the attractors and basins for the ten thousand networks within a reasonable time, 26 nodes ($2^{26}$=67,108,864 states) are the maximum network size we can handle in our computing environment. Table 1 shows information about the biological networks collected from the *Cell Collective* public platform for modeling biological networks. In the table, the networks are sorted by their size.

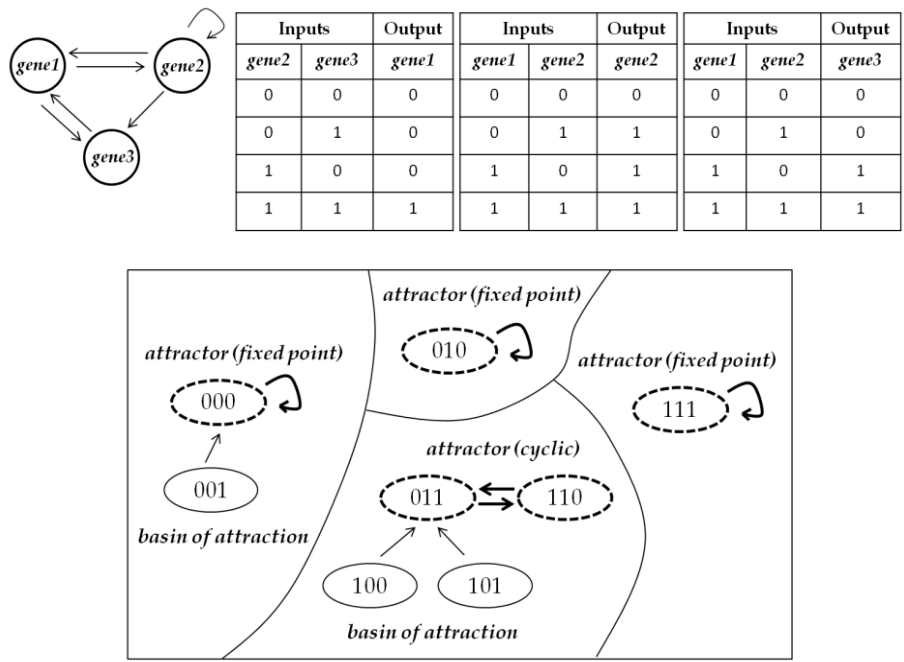

**Figure 1.** An example RBN with *N*=3 and *K*=2 and its state transition diagram. The topology is randomly generated and Boolean functions are randomly assigned to each node. The state transition diagram is composed of $2^3$=8 state configurations from 000 to 111 and transitions among them. In the state transition diagram, attractors are the configurations with bold dashed lines, and basins of attraction are the configurations except for the attractors.

**Table 1.** Different Boolean network models of 37 biological systems used for simulations.

| Biological network [1] | # of nodes | # of links | Ref. |
|---|---|---|---|
| #1. Cortical area development (cortical) | 5 | 14 | [36] |
| #2. Cell cycle transcription by coupled CDK and network oscillators (cycle-cdk) | 9 | 19 | [37] |
| #3. Mammalian cell cycle (core-cell-cycle) | 10 | 35 | [38] |
| #4. Toll pathway of drosophila signaling pathway (toll-drosophila) | 11 | 11 | [39] |
| #5. Metabolic interactions in the gut microbiome (metabolic) | 12 | 30 | [40] |
| #6. Regulation of the L-arabinose operon of Escherichia coli (l-arabinose-operon) | 13 | 18 | [41] |
| #7. Lac operon (lac-operon-bistability) | 13 | 22 | [42] |
| #8. Arabidopsis thaliana cell-cycle (arabidopsis) | 14 | 66 | [43] |
| #9. Fanconi anemia and checkpoint recovery (anemia) | 15 | 66 | [44] |
| #10. Cardiac development (cardiac) | 15 | 38 | [45] |
| #11. BT474 breast cell line short-term ErbB network (bt474-ErbB) | 16 | 46 | [44, 46] |
| #12. SKBR3 breast cell line short-term ErbB network (skbr3-short) | 16 | 41 | [46] |
| #13. Neurotransmitter signaling pathway (neurotransmitter) | 16 | 22 | [47] |
| #14. HCC1954 breast cell line short-term ErbB network (hcc1954-ErbB) | 16 | 46 | [46] |
| #15. Body segmentation in drosophila (body-drosophila) | 17 | 29 | [48] |
| #16. CD4+ T cell differentiation and plasticity (cd4) | 18 | 78 | [49] |
| #17. Budding yeast cell cycle (budding-yeast) | 18 | 59 | [50] |
| #18. T-LGL survival network (t-lgl-survival) | 18 | 43 | [51] |
| #19. VEGF pathway of drosophila signaling pathway (vegf-drosophila) | 18 | 18 | [39] |
| #20. Oxidative stress pathway (oxidative-stress) | 19 | 32 | [49, 52] |
| #21. Human gonadal sex determination (gonadal) | 19 | 79 | [53] |
| #22. Mammalian cell-cycle (mammalian) | 20 | 51 | [52, 54] |
| #23. Budding yeast cell cycle (yeast-cycle) | 20 | 42 | [55] |
| #24. B cell differentiation (b-cell) | 22 | 39 | [56] |
| #25. Iron acquisition and oxidative stress response in aspergillus fumigatus (aspergillus-fumigatus) | 22 | 38 | [57] |
| #26. FGF pathway of drosophila signaling pathways (fgf-drosophila) | 23 | 24 | [39] |
| #27. T cell differentiation (t-cell-differentiation) | 23 | 34 | [58] |
| #28. Aurora kinase A in neuroblastoma (aurka) | 23 | 43 | [59] |
| #29. Processing of Spz Network from the drosophila signaling pathway (spz-drosophila) | 24 | 28 | [39] |
| #30. TOL regulatory network (tol) | 24 | 48 | [60] |
| #31. HH pathway of drosophila signaling pathways (hh-drosophila) | 24 | 32 | [39] |
| #32. HCC1954 breast cell line long-term ErbB network (hcc1954) | 25 | 70 | [46] |
| #33. SKBR3 breast cell line long-term ErbB network (skbr3-long) | 25 | 81 | [46] |
| #34. BT474 breast cell line long-term ErbB network (bt474) | 25 | 70 | [46] |
| #35. Wg pathway of drosophila signaling pathways (wg-drosophila) | 26 | 29 | [39] |
| #36. Trichostrongylus retortaeformis (trichostrongylus) | 26 | 58 | [61] |
| #37. Pro-inflammatory tumor microenvironment in acute lymphoblastic leukemia (leukemia) | 26 | 81 | [62] |

[1] Data was obtained from *Cell Collective* (https://research.cellcollective.org/?dashboard=true#).

*2.2. Mutations and Classification of Robustness & Evolvability*

We introduce four types of random mutations to each biological network to study their robustness and evolvability: we add, delete one regulatory link, change the position of a link in the network, or flip one state (*i.e.*, 0 changes into 1, and 1 changes into 0) in the output of Boolean functions assigned to each node [63-65]. These four types of mutations are distributed as equally as possible. We then measure properties of different mutants without repetitions and compare them with the original networks.

Comparing the attractors between the original and mutated networks, we classify the properties of biological networks [66]. Our classification is extended from the definition of robust and evolvable network in Aldana *et al.*'s work [14]. They added perturbations to a network structure and then observed attractors between the original and perturbed networks. Under the assumption that all attractors represent essential cell types or cell functions, they considered the preservation of attractors *robustness* and the emergence of new attractors *evolvability*.

Based on their concept, we divide the properties of robustness and evolvability into four classes: *not robust & not evolvable*, *not robust & evolvable*, *robust & not evolvable*, and *robust & evolvable* (Figure 2). For example, if an original network has a set of attractors $A = \{a_1, a_2, a_3\}$.

- *not robust & not evolvable*: The mutated network does not have exactly the same attractors as the attractors of the original network and further does not produce any new attractors. It is the case where the mutated network has a set of attractors $A' = \{a_1, a_2\}$ ($A \supset A'$).
- *not robust & evolvable*: The mutated network does not fully have original attractors but creates new attractors. It is the case where the mutated network has a set of attractors $A' = \{a_1, a_4\}$ ($A \neq A'$ and $A \not\subset A'$ and $A' \not\subset A$).
- *robust & not evolvable*: The mutated network maintains the original attractors but does not generate any new attractors. It is the case where the mutated network has a set of attractors $A' = \{a_1, a_2, a_3\}$ ($A = A'$).
- *robust & evolvable*: The mutated network has the same attractors as the original one and simultaneously produces new attractors. It is the case where the mutated network has a set of attractors $A' = \{a_1, a_2, a_3, a_4, a_5\}$ ($A \subset A'$).

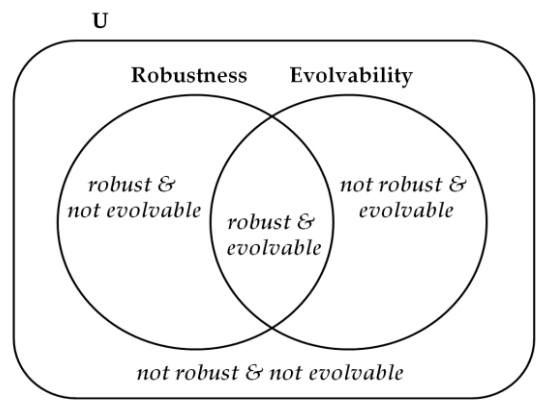

**Figure 2.** The schematic diagram of the four classes on robustness and evolvability. Depending on the change of attractors between original and mutated networks, the network is certainly classified into one class among *not robust & not evolvable*, *not robust & evolvable*, *robust & not evolvable*, and *robust & evolvable*.

In our simulations, we independently add an internal perturbation 1,000 times to each network, and then get 1,000 perturbed networks per biological network. Next, we classify the properties of all the perturbed networks into the above four classes. Since one biological network can be *not robust &*

*not evolvable*, *not robust & evolvable*, *robust & not evolvable*, or *robust & evolvable* against the 1,000 perturbations, we can get the percentage frequency distribution of the four classes per network.

*2.3. Antifragility in Boolean Networks*

Antifragility was defined by Taleb [22]. Antifragility is the property of a system that improves when exposed to perturbations. We recently developed a measure to assess antifragility in Boolean networks [23, 24]. In this study, we employed this antifragility measure. One advantage of this measure is that its computation time increases linearly with *N* (S2 in Supplementary Material).

On the assumption that antifragility indicating responses to external perturbations might be helpful to predict responses to internal perturbations (*i.e.*, mutations), we use antifragility as a predictor to estimate the robustness and evolvability of biological networks against mutations. To understand our antifragility measure, we first have to explain how we measure complexity and external perturbations.

2.3.1. Complexity of Boolean networks

Complexity can be seen as a balance between change and regularity [26]. In terms of information, *change* means that information becomes different, while *regularity* means that information is preserved. In biology, a balance between the change and preservation of genetic information enables biological systems to have flexibility and stability by which biological systems robustly adapt to their environment.

We defined the change and regularity as emergence and self-organization respectively and developed measures to quantify them. Also, using the two metrics, we presented a measure to evaluate complexity [67, 68]. Complexity (*C*) is calculated from emergence (*E*) and self-organization (*S*) as follows:

$$C = 4 \times E \times S, \qquad (1)$$

where coefficient 4 is added to normalize the values of *C* to the range of [0,1] ($0 \leq C \leq 1$). Because *S* can be regarded as the complement of *E* [67, 68], equation (1) is reformulated by the following equation:

$$C = 4 \times E \times (1 - E), \qquad (2)$$

In a Boolean network which consists of *N* nodes, *E* ($0 \leq E \leq 1$) is measured as the average of emergence values for all nodes, where the emergence of each node is computed through Shannon's information entropy:

$$E = \frac{\sum_{i=1}^{N} E_i}{N} = \frac{\sum_{i=1}^{N} -(p_0^{(i)} log_2 p_0^{(i)} + p_1^{(i)} log_2 p_1^{(i)})}{N}, \qquad (3)$$

where $p_0^{(i)}$ ($p_1^{(i)}$) is the ratio of how many 0s (1s) are expressed to *T* state transitions for node *i* (*i.e.*, $p_0^{(i)} + p_1^{(i)} = 1$). Regarding *T*, we obtain $p_0^{(i)}$ ($p_1^{(i)}$) on state transitions not from 1 to *T* but from *T*+1 to 2*T* so that we can exclude as many transient states as possible. This enables us to get more stable values of $p_0^{(i)}$ ($p_1^{(i)}$) by focusing on states closer to or belonging to attractors (S1 in Supplementary Material). Figure 3 shows an example of computing *E* in a Boolean network.

In equation (2), *C* has the maximum value of 1 when *E* is 0.5 [68, 69]. For example, under the condition that all the nodes of a Boolean network have the same emergence values, when $p_0$ or $p_1$ is around 0.89, average *E* becomes 0.5. In contrast, *C* has the minimum value of 0 for a node when *E* is 1 (constant change) or 0 (no change). Under the condition that all the nodes have the same emergence values, when $p_0$ and $p_1$ are 0.5 or when $p_0$ or $p_1$ is 1, average *E* of a network becomes 1 or 0, respectively. As seen in the examples, complexity is determined by how 1s and 0s are distributed during state transitions. The distribution of 1s and 0s at each node represents how node states are altered and kept. Thus, the complexity of Boolean networks measures a degree of the balance ($E$ =0.5) between change ($E$ =1) and regularity (E=0) of gene expression information.

2.3.2. External perturbations to Boolean networks

We consider flipping the node states of the networks as *external* perturbations so as to measure antifragility [23, 24] (as opposed to the *internal* perturbations, *i.e.*, mutations, described above). The degree of external perturbations ($\Delta x$) is quantified by the following equation:

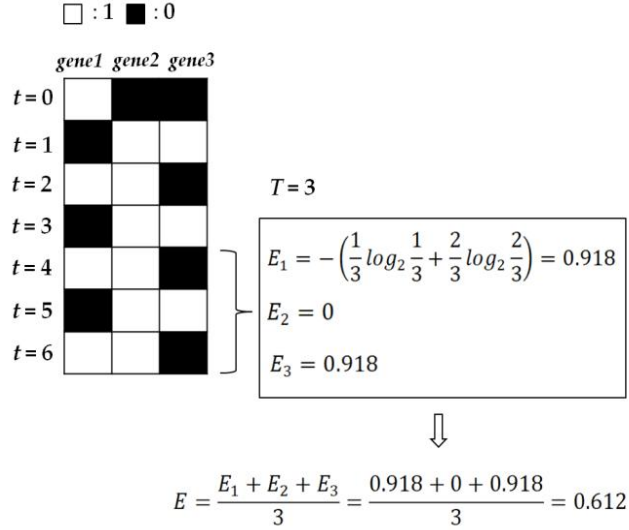

$$\Delta x = \frac{X \times (\frac{T}{O})}{N \times T} = \frac{X}{N \times O},\qquad(4)$$

**Figure 3.** An example showing how to calculate the emergence of each node and the average, $E$. Because the averaged emergence $E$ (of the network) is 0.612, the complexity $C$ is $4 \times 0.612 \times (1 - 0.612) \cong 0.95$. With initial configuration 100, the state transitions were obtained from $t$=0 to $t$=6 in the example RBN of Figure 1.

where $X$ is the number of nodes randomly chosen to be perturbed, $T$ is the simulation time for state transitions, and $O$ is the perturbation frequency which determines by how many time steps are executed between perturbations. $X$ nodes are randomly chosen and perturbed every time step of $O$. For instance, if $N$=10, $X$=4, $T$=5, and $O$=1, we randomly choose 4 nodes ($X$=4) in a network with 10 nodes ($N$=10) and then change the states of the selected nodes. We repeat this perturbation every time step ($O$=1) during 2×5=10 time steps ($T$=5) Here, perturbing the network during not 5 but 10 timesteps is because we are interested in state transitions from $T$+1 to $2T$ as mentioned in 2.3.1.

$\Delta x$ has the interval [0,1] ($0 \leq \Delta x \leq 1$). This term is used to adjust the influence of network size on antifragility. In the simulations, the parameters are set to $X$=[1, 2, ..., $N$], $T$=200, and $O$=1. The values of $T$ and $O$ were determined based on our previous research [23]: if $T$ is large enough, the shape of the antifragility curve is consistently obtained over $T$ (with $T$ increasing). In other words, antifragility curves quickly converge with an increasing $T$. For Boolean networks which have less than 100 nodes, $T$=200 was enough to measure antifragility. $O$=1 showed the most distinctive difference of antifragility between networks. As with other variables related to our antifragility measure, the time required to perform these calculations increases linearly with $N$.

2.3.3. Antifragility of Boolean networks

Our antifragility measure ($\oint$) is composed of two terms: the difference of "satisfaction" before and after external perturbations ($\Delta\sigma$) and the degree of external perturbations ($\Delta x$) [23, 24]. The equation is as follows:

$$\oint = -\Delta\sigma \times \Delta x, \qquad (5)$$

Here the degree of external perturbations ($\Delta x$) was explained in 2.3.2. To obtain the difference of *satisfaction* before and after external perturbations ($\Delta\sigma$), the concept of satisfaction ($\sigma$) should be explained. $\sigma$ represents how much the "goal" of agents has been attained [70]. The agents and goal can be defined differently depending on systems and observers. In Boolean networks, each node can be regarded as an agent. Their goal can be arbitrarily defined as achieving high complexity.

For equation (5), $\Delta\sigma$ is calculated by the following equation:

$$\Delta\sigma = C - C_0, \qquad (6)$$

where $C_0$ and $C$ are the complexities before and after external perturbations, respectively. $\Delta\sigma$ has values in the range [-1,1] ($-1 \leq \Delta\sigma \leq 1$) as $C_0$ and $C$ have the interval [0,1]. Thus, if $\Delta\sigma$ is positive ($C \geq C_0$), it means that complexity was increased by the external perturbations. That is, we define "benefiting from perturbations" as increasing their complexity. If this is the case, the Boolean network can be considered *antifragile*. If $\Delta\sigma$ is negative ($C \leq C_0$), it indicates that complexity decreases by external perturbations. It can be seen as *fragile*. If $\Delta\sigma$ is zero ($C = C_0$), complexity is maintained against external perturbations. Thus, the network can be regarded as *robust*.

To calculate $C_0$ and $C$, when the state transitions of the original network and perturbed one are computed, the same initial states are used at *t*=0. Depending on the initial states, because complexity can be different, $C_0$ and $C$ are calculated as the average of the complexity values acquired from a large number of initial conditions that are randomly chosen. This yields a more stable value of $\oint$ as a system property. In our simulations, the number of initial states (*s*) is set to 1,000.

*2.4. Property Classification with a Convolutional Neural Network*

We tried several classification methods, including multinomial logistic regression (46.33% accuracy), and a convolutional neural network (CCN) was the most effective.

2.4.1. Input and output in a CNN

We use a CNN to classify the properties of the biological networks into the four classes: *not robust & not evolvable*, *not robust & evolvable*, *robust & not evolvable*, and *robust & evolvable*. In our CNN model, the input consists of the differences of antifragility between the original and mutated networks, and the output consists of the properties classified into the four classes. As we use 37 biological networks and add the internal perturbations 1,000 times to each network, we have 37×1,000=37,000 quantities of the differences of antifragility and the classified properties, respectively.

We explain how to obtain the input, taking the *Mammalian cell cycle* network with 20 nodes and its mutated network as examples. Figures 4(a) and 4(b) show the antifragility of the two networks depending on the number of perturbed nodes *X*. We interpolate to get 30 data points for each network (independently of *N*) and subtract the antifragility value of the mutated network from that of the original network at each data point. Then, we get 30 difference values in the normalized range [1/*N*, 1 (=*N*/*N*)] (Figure 4(c)). In this way, for all the networks with the different number of nodes, we can get 37,000 input elements in which one element is composed of 30 data points. This process allows antifragility of all the networks to have the same size of 30×1. This shape of input is necessary to use CNNs which require images with the same width and height as input.

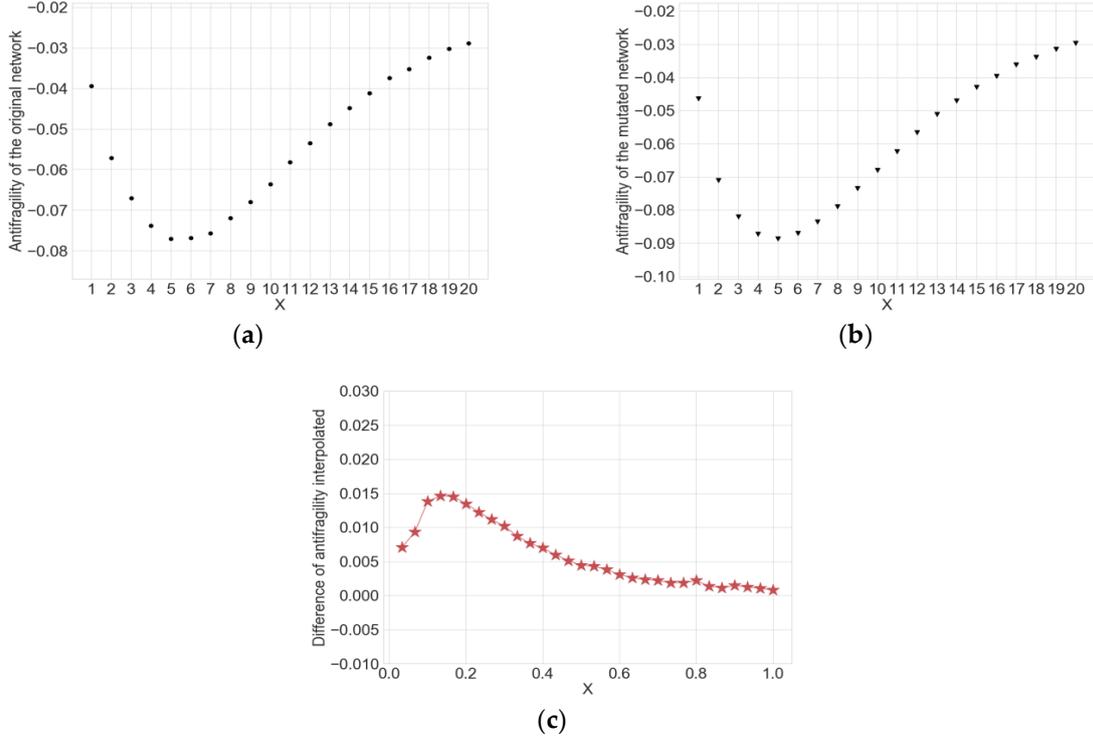

**Figure 4.** Data points related to antifragility of the *mammalian cell cycle* network with $N$=20. In the simulations, the parameters were set to perturbed node size $X$=[1, 2, ..., 20], simulation time for state transitions $T$=200, perturbation frequency $O$=1: (**a**) 20 data points on antifragility of the original network; (**b**) 20 data points on antifragility of the mutated network; (**c**) 30 data points on the differences of antifragility estimated through interpolation in the normalized range.

2.4.2. Nested k-fold cross-validation

We carry out *nested k-fold cross validation* to select a model and evaluate the model performance. Figure 5 briefly illustrates the process of the cross-validation. We split the data with the 37,000 quantities into training data (*i.e.*, 37,000×0.75 = 27,750) and test data (*i.e.*, 37,000×0.25 = 9,250) with the ratio of 75 to 25 and set $k$=4 (Outer loop A in Figure 5). However, the data about the properties labeled with the four classes is imbalanced, as the four classes are not equally distributed. Thus, we have a multiclass classification problem with imbalanced data. We balance the training data using an oversampling technique. We get balanced data that make up of the same number of samples per class. Then, we divide the balanced data into training data and validation data with the ratio of 67 to 33 (Inner loop in Figure 5). From $k$=1 to $k$=4, we train our model with the balanced data and test it with the original imbalanced data for each fold (Outer loop B in Figure 5).

We have three hyperparameters: batch size={32, 64, 128}, data balancing={SMOTE, ADASYN}, and CNN architecture={simple, complex}. For the data balancing, we try SMOTE (Synthetic Minority Over-sampling Technique) and ADASYN (Adaptive Synthetic Sampling Approach), which are oversampling techniques generating synthetic samples from the minority class [71, 72]. Regarding the architecture, we try a simple CNN model and a complex one. CNN is usually composed of the convolutional layer (Conv) with a rectified linear unit (ReLU) activation function, pooling layer (Pool), flattening layer (Flat), fully connected layer (FC), and output layer (Out). The layers are arranged in the order of Conv+ReLU-Pool-Flat-FC-Out. Depending on how many Conv+ReLU-Pool layers are stacked, CNNs become deeper neural networks. The deeper CNNs can handle higher resolution images. In our study, the simple CNN has Conv-Conv+ReLU-Pool, and the complex CNN has Conv+ReLU-Conv+ReLU-Pool-Conv+ReLU-Conv+ReLU-Pool (Figure 6).

- **Inner loop:** We set the values of hyperparameters and fit the model parameters. The model is trained and validated with the balanced data. We use AUC (*i.e.,* area under the ROC curve) as a criterion to find the best hyperparameter set. We have 12 hyperparameter sets in total (Table 2 and Figure 6). For the three rows of the inner loop derived from the first fold in Figure 5, we calculate AUC = {AUC(hyp1), AUC(hyp2), …, AUC(hyp12)} in the validation of each row, and then get averages over the three rows (i.e., $AUC^{avg.}$ = {$AUC(hyp1)^{avg.}$, $AUC(hyp2)^{avg.}$, …, $AUC(hyp12)^{avg.}$}). Repeating the process from the second fold to the fourth one, we get their means for the four folds (i.e., $AUC^{final}$ = {$AUC(hyp1)^{final}$, $AUC(hyp2)^{final}$, …, $AUC(hyp12)^{final}$}). Finally, we selected hyp 2 as the best hyperparameter set from the values of $AUC^{final}$ in Table 2.

- **Outer loop:** We train the model which has the optimal parameters determined from the inner loop. The model is trained with the balanced data. Table 3 and Table 4 show the accuracy of the training and validation dataset, respectively. After training the model, we test it with the original imbalanced data (Outer loop B in Figure 5). To evaluate the model performance, we get a test accuracy, a confusion matrix, and precision-recall (PR) curves. They are averages over the four folds.

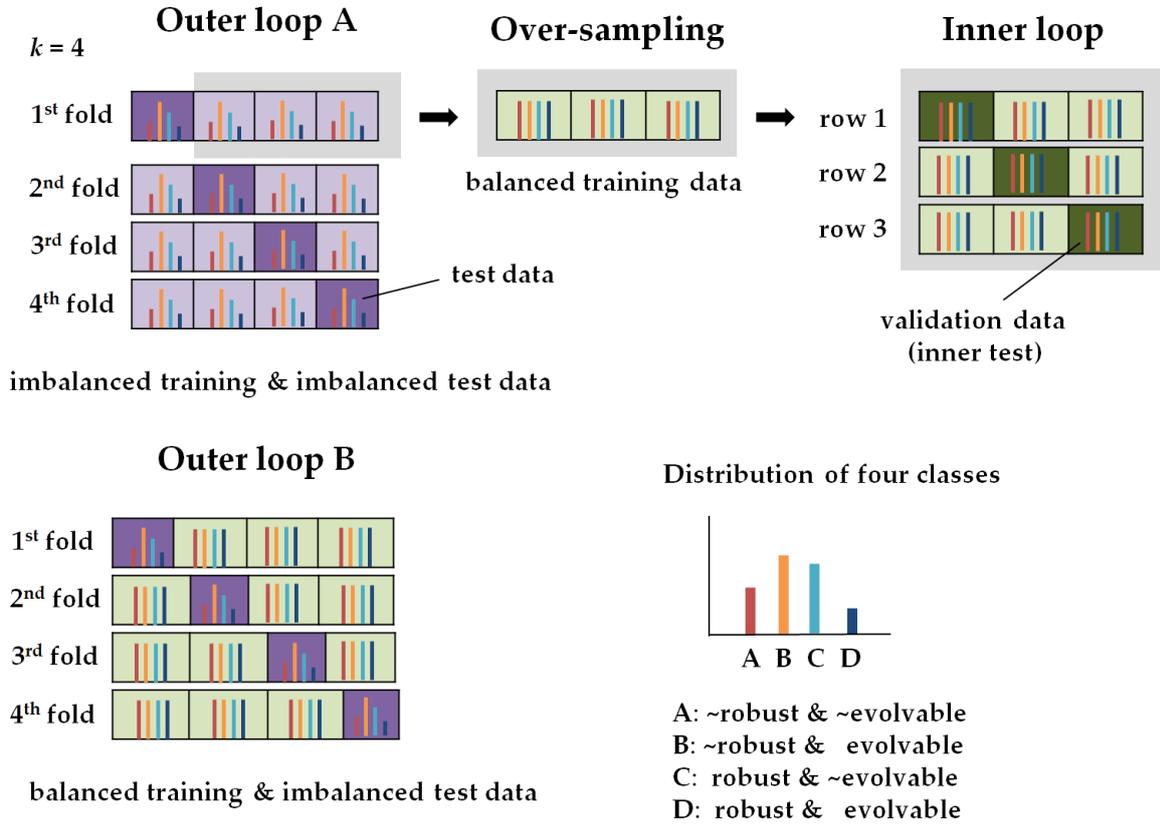

**Figure 5.** The illustration for the processes of the nested k-fold cross-validation (k=4). In the inner loop, the values of the hyperparameters are set and the model parameters are fitted. In the outer loop, the model performance is evaluated.

Table 2. 12 hyperparameter sets for the simulations in the inner loop.

| Set | Epoch | Batch size | Balancing | Architecture | AUC$^{final}$ |
|---|---|---|---|---|---|
| hyp 1 | 128 | 32 | SMOTE | simple | 0.8265 |
| hyp 2 | 128 | 64 | SMOTE | simple | 0.8295 |
| hyp3 | 128 | 128 | SMOTE | simple | 0.8251 |
| hyp4 | 128 | 32 | SMOTE | complex | 0.8075 |
| hyp5 | 128 | 64 | SMOTE | complex | 0.8101 |
| hyp6 | 128 | 128 | SMOTE | complex | 0.8056 |
| hyp7 | 128 | 32 | ADASYN | simple | 0.8151 |
| hyp8 | 128 | 64 | ADASYN | simple | 0.8124 |
| hyp9 | 128 | 128 | ADASYN | simple | 0.8137 |
| hyp10 | 128 | 32 | ADASYN | complex | 0.7922 |
| hyp11 | 128 | 64 | ADASYN | complex | 0.7931 |
| hyp12 | 128 | 128 | ADASYN | complex | 0.7941 |

**Table 3.** Training accuracy.

| | *row 1* | *row 2* | *row 3* |
|---|---|---|---|
| *1st fold* | 0.7266 | 0.7288 | 0.6929 |
| *2nd fold* | 0.7410 | 0.7362 | 0.7433 |
| *3rd fold* | 0.7324 | 0.7085 | 0.7134 |
| *4th fold* | 0.6889 | 0.7030 | 0.7212 |
| | | | avg.=0.7197 |

**Table 4.** Validation accuracy.

| | *row 1* | *row 2* | *row 3* |
|---|---|---|---|
| *1st fold* | 0.5688 | 0.5802 | 0.5815 |
| *2nd fold* | 0.5965 | 0.5661 | 0.6003 |
| *3rd fold* | 0.6078 | 0.5828 | 0.5547 |
| *4th fold* | 0.5682 | 0.5742 | 0.5630 |
| | | | avg.=0.5787 |

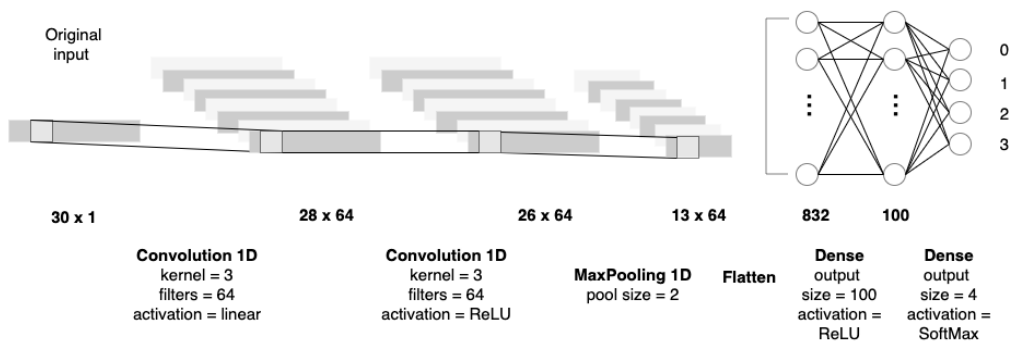

(a)

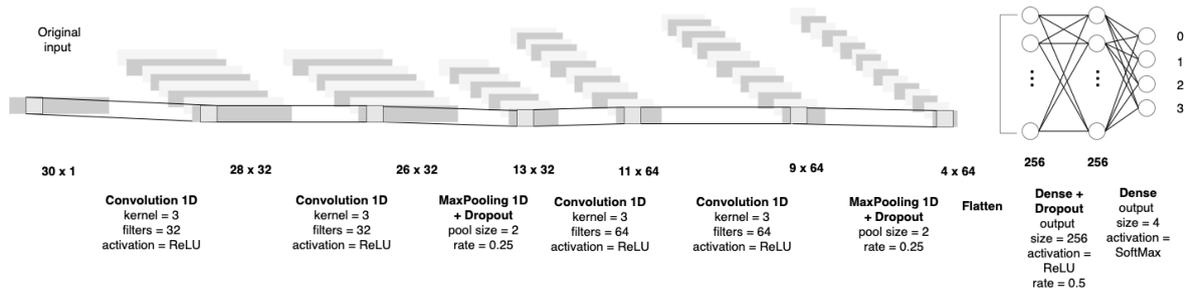

**(b)**

**Figure 6.** CNN architectures for simulations. (a) Our simple CNN model has two convolution layers, and one pooling layer. (b) Our complex CNN model has four convolution layers and two pooling layers.

## 3. Results

### 3.1. Attractors and Basins of Attraction of Biological Networks

We explored the attractors and basins of attraction to find the structural features of the state space in the Boolean network models of the 37 biological systems. Specifically, we measured the number of attractors, the average length of attractors, and the normalized basin entropy. In the case of the normalized basin entropy, it was calculated it by dividing Krawitz, *et al.*'s measure [73] by the number of nodes (*i.e.*, $H = -\sum_\rho p_\rho \log_2 p_\rho /N$ where $p_\rho$ is the basin size of attractor $\rho$ divided by state space size $2^N$, $\sum_\rho p_\rho = 1$). This is normalized between 0 and 1. As mentioned in the introduction, the attractors can represent cell types or functions [15-17]. Hence, from a biological viewpoint, the number of attractors can be interpreted as the number of cell functions, and the average length of attractors can be regarded as the time that it takes cell functions to be conducted. The normalized basin entropy can be seen as the versatility of cell functions because it provides information about the distribution of basin sizes. The more even the basin sizes of attractors are in the state space, the larger the value of normalized entropy is. In the case that all basins of attraction have similar size, when initial states are changed by noise, they are more likely to converge different attractors by jumping from one basin to another.

Figure 7(a) shows the distribution of the number of attractors in log scale. It is clear that various biological systems have different number of cell functions. Figure 7(b) displays the distribution of the average length of attractors. Overall, more than half of the networks have only fixed-point attractors, and the rest of them have relatively short cyclic attractors when their state space sizes are considered. The values varied between 1 and 11, indicating that each biological system spends different time carrying out its cellular function(s). Figure 7(c) presents the distribution of the normalized basin entropy. The values are diversely distributed between 0 and 0.541. This indicates that some biological systems perform only a few cell functions dominantly, while others commonly change the cell functions among many ones.

All three measures have large value variations, so it is difficult to find common structural features of the state space. These biological systems are specialized for their different cell functions, which may cause such variations. From the variations, we can see that there exist biological networks that have a broad variety of number of attractors, attractor lengths, and basin distributions. For these networks, it is computationally difficult to study the robustness and evolvability of biological systems by comparing the attractors before and after perturbations. It is even harder for larger biological networks because the number of attractors, the attractor length, and the basin size will increase as the number of nodes grows. Therefore, it is necessary to estimate the robustness and evolvability of biological networks without measuring attractors exhaustively.

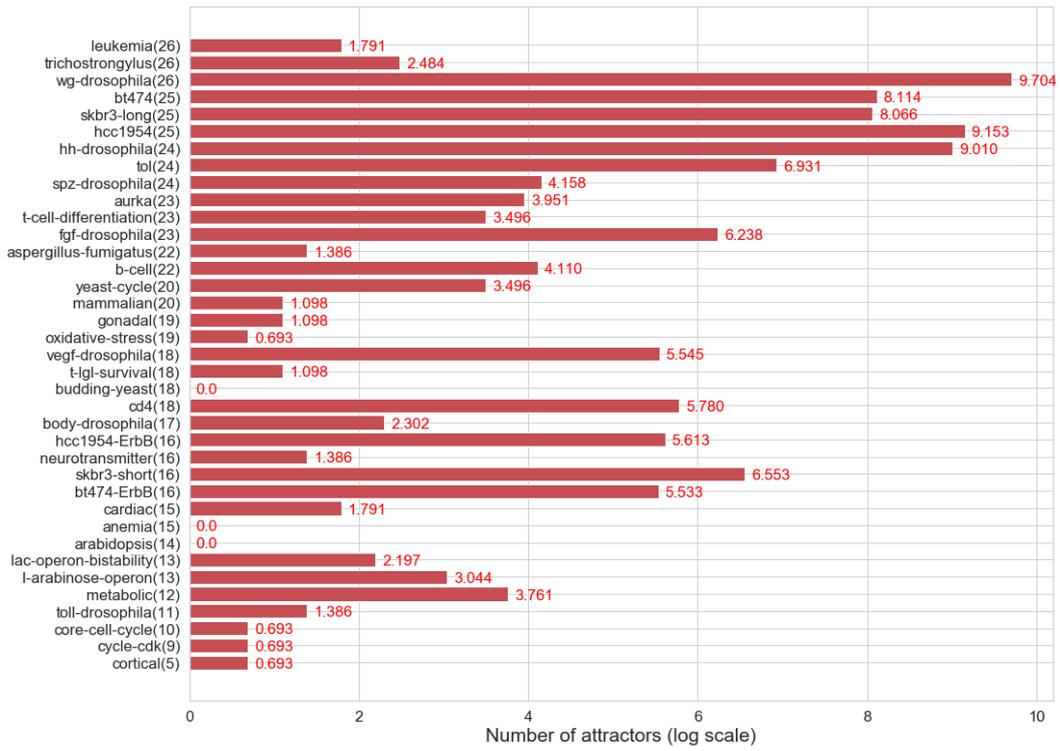

(a)

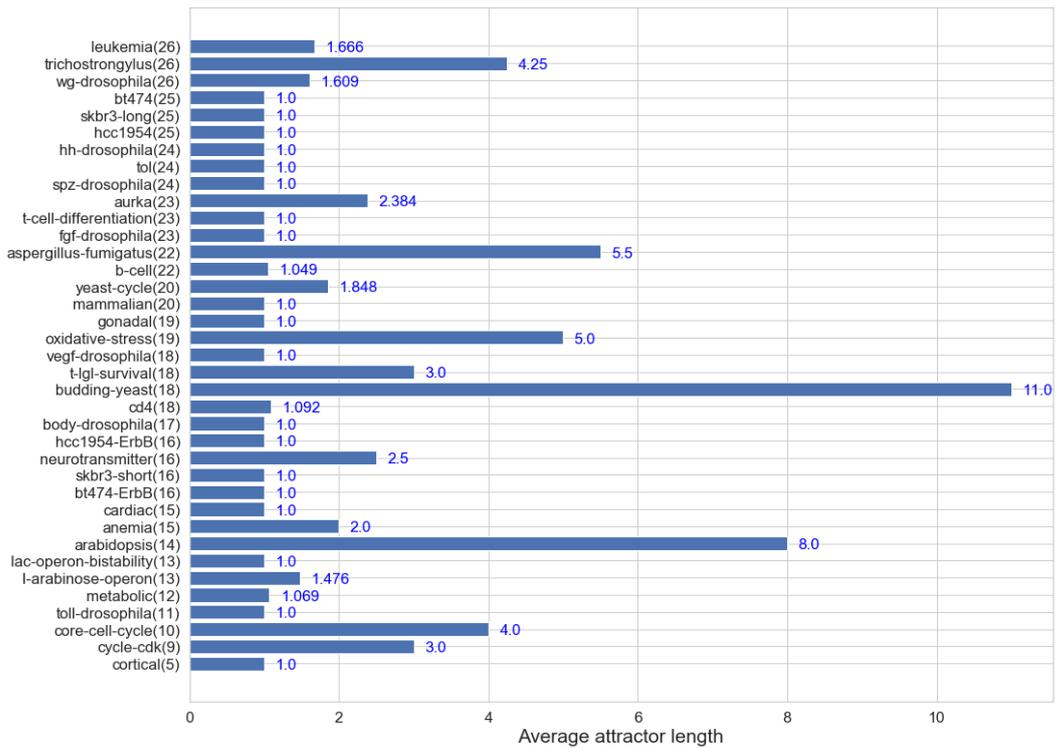

(b)

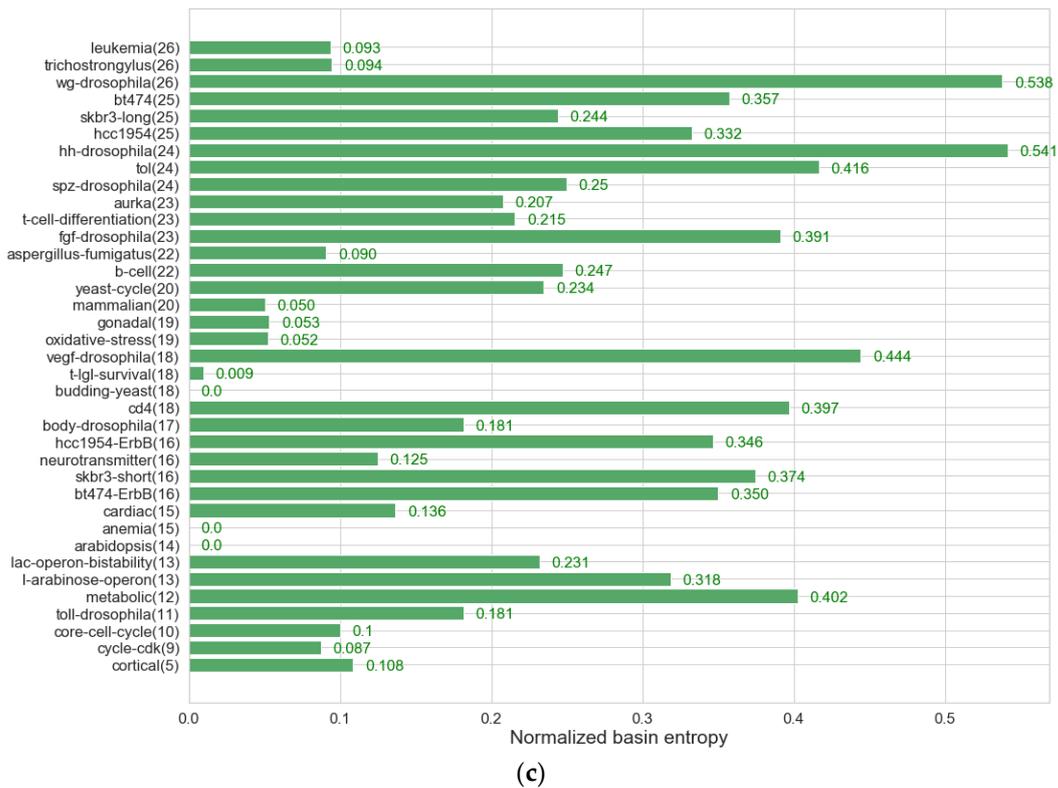

(c)

**Figure 7.** Attractors and basins of attraction of the 37 biological networks. The number in the parenthesis on the y-axis points out the number of nodes of the network: (**a**) The number of attractors. The values took the natural logarithm ($e \cong 2.718$) so 0 means that the biological network has a single attractor.; (**b**) The average length of attractors; (**c**) Normalized basin entropy. It has the range [0,1]. The more even the basin sizes of attractors are in the state space, the larger the value normalized entropy will have.

*3.2. Distribution of the Four Classes on Robustness & Evolvability of Biological Networks*

We classified the robustness and evolvability of the 37 biological networks into the four classes (*not robust & not evolvable*, *not robust & evolvable*, *robust & not evolvable*, *robust & evolvable*) to investigate how the biological networks respond to mutations. Figure 8 shows the percentage frequency distribution of the four classes. For each biological network, we added a different internal perturbation 1,000 times, and thus acquired the percentage frequency from 1,000 different mutated networks.

As seen in Figure 8, every biological network has three, or four classes. Overall, the class of *robust & evolvable* takes up the lowest percentage. The classes of *not robust & evolvable* and *robust & not evolvable* account for the majority of the percentage. The only class that is not present in all networks is *not robust & not evolvable*. These findings propose that a biological system can display different behaviors of robustness and evolvability against mutations. Furthermore, it is known that biological systems are robust and evolvable [1-6], which might be able to explained from not the behavior of *robust & evolvable* but the two dynamical behaviors of *not robust & evolvable* and *robust & not evolvable*. In other words, it is highly likely that mutations will lead to *robust* OR *evolvable* networks. It can also be argued that a more complex mechanism is required for being at the same time *robust & evolvable*, so this could explain why not many networks exhibit both properties frequently.

In addition, such distribution of the dynamical responses suggests evolutionary profiles of what the environments given to the biological systems were like. *not robust & evolvable* makes up the highest percentage frequency of the four classes in the distribution, which implies that many biological systems might have been mainly exposed to drastic environmental changes hard to keep existing cell functions, and thus evolved towards preferring producing brand new cell functions to adapt to the new environments.

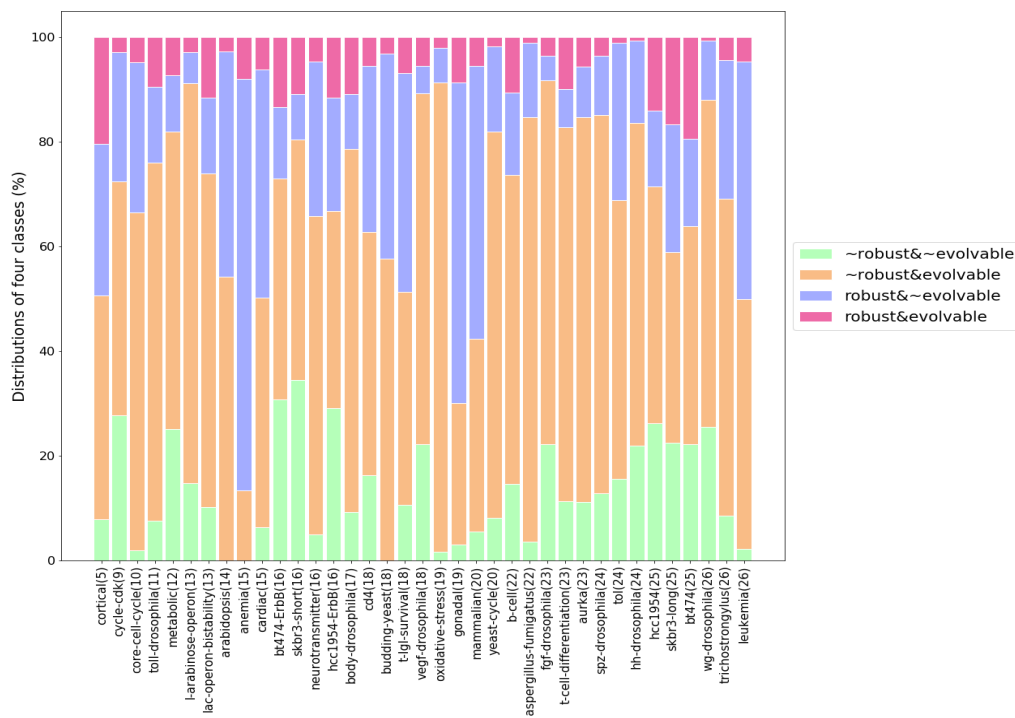

**Figure 8.** Percentage frequency distribution of the four classes on robustness and evolvability for the 37 biological networks. A different internal perturbation was added to each network 1,000 times, so 1,000 different mutated networks were generated per biological network. The perturbed networks were classified into *not robust & not evolvable*, *not robust & evolvable*, *robust & not evolvable*, or *robust & evolvable*.

*3.3. Association Between Mutation Type and Robustness & Evolvability*

We computed Cramer's V to measure the strength of association between the mutation type and the robustness and evolvability. We had the four mutation types: adding a link (*add*), deleting a link (*delete*), changing the position of a link (*change*), and flipping one state in a Boolean function (*flip*). We randomly chose one mutation type and then added it to the biological network. This process was repeated 1,000 times per biological network, so we obtained 37,000 pairs (mutation type, property class) for the 37 biological networks. Table 5 is a contingency table displaying the frequency distribution of the four classes depending on the mutation type. From the table, we got Cramer's V=0.2292. Cramer's V takes values from 0 to 1. The closer to zero, the weaker association between the variables. Hence, we found that there is a weak association between the mutation type and the property class. This indicates that the type of genetic mutations do not have a strong effect on determining the robustness and evolvability of biological systems.

**Table 5.** A contingency table for the four classes depending on the mutation type.

|  | *add* | *delete* | *change* | *flip* |
| --- | --- | --- | --- | --- |
| *not robust & not evolvable* | 2180 | 549 | 1731 | 532 |
| *not robust & evolvable* | 11443 | 1204 | 6067 | 1889 |
| *robust & not evolvable* | 2380 | 642 | 1873 | 3893 |
| *robust & evolvable* | 958 | 241 | 754 | 664 |

*3.4. Prediction of Robustness & Evolvability Using Antifragility*

We got a test accuracy, confusion matrix, and precision-recall curves for the test data to evaluate the performance of our CNN model. We preferred using precision-recall (PR) curves to receiver operating characteristic (ROC) curves because we tested the model with the imbalanced data [74].

Table 6 presents a test accuracy. The overall accuracy is 0.5845. Figure 9(a) shows a normalized confusion matrix. In the confusion matrix, the x-axis refers to an instance of the predicted classes, and the y-axis represents an instance of the actual classes. The values of the diagonal elements mean the probability of correctly predicted classes. For each class, *not robust & not evolvable* is 0.6150, *not robust & evolvable* is 0.6632, *robust & not evolvable* is 0.5657 and *robust & evolvable* is 0.4108. From the accuracy and confusion matrix, we can see that our model has a satisfactory performance for the classification of robustness and evolvability.

Figure 9(b) exhibits a micro averaged PR curve for the four classes. We computed the micro average globally not distinguishing the elements between different classes, which is usually preferable for imbalanced classes. In our PR curve, the average precision (AP) is 0.54. This score refers to the area under the PR curve. To compute AP, the function of *average_precision_score* was used in the scikit-learn library of Python[1]. The large AP means high precision and high recall. High precision is related to a low false positive rate (type I error  α) and high recall is related to a low false negative rate (type II error  β).

Overall, our model has a good performance but there are differences for each class. Figure 10 shows PR curves per class. *not robust & not evolvable* has AP=0.52 (Figure 10(a)), *not robust & evolvable* has AP=0.64 (Figure 10(b)), *robust & not evolvable* has AP=0.81 (Figure 10(c)), and *robust & evolvable* has AP=0.17 (Figure 10(d)). The three classes of *not robust & not evolvable*, *not robust & evolvable*, and *robust & not evolvable* have larger AP values when compared to the class of *robust & evolvable*. It indicates that our classifier not only detects a majority of the positive results for the three classes but also classify them correctly. Meanwhile, *robust & evolvable* has the smallest AP value. It might be a result from the scarcity of data on *robust & evolvable*, as seen in Figure 8, and/or a high variance of the differences of antifragility classified into *robust & evolvable*. The two things might have made it hard for our CNN model to find the feature map of *robust & evolvable*.

**Table 6.** Test accuracy.

|  |  |
|---|---|
| *1st fold* | 0.5794 |
| *2nd fold* | 0.5676 |
| *3rd fold* | 0.5897 |
| *4th fold* | 0.6015 |
|  | avg.=0.5845 |

Table 7 shows the AP values between our model and random classifiers. In PR curves, AP of random classifiers is 0.5 only when there are two classes and their distributions are balanced. For binary classification of balanced and imbalanced datasets, AP of random classifiers is defined as positive/(positive + negative)  [63, 74]. Our test data is imbalanced, so we computed the AP values of random classifiers based on the ratio of positive and negative. Because output is binarized when the PR curve and AP are extended to multi-class classification, calculating the AP values of our model and random classifiers is regarded as binary classification for each class. Table 7 demonstrates that all the AP values for the four classes in our model are larger than those of random classifiers. Although the class of *robust & evolvable* has relatively smaller AP value than the other classes, our model has better performance to classify all the classes than random classifiers.

---

[1] https://scikit-learn.org/stable/auto_examples/model_selection/plot_precision_recall.html#sphx-glr-auto-examples-model-selection-plot-precision-recall-py

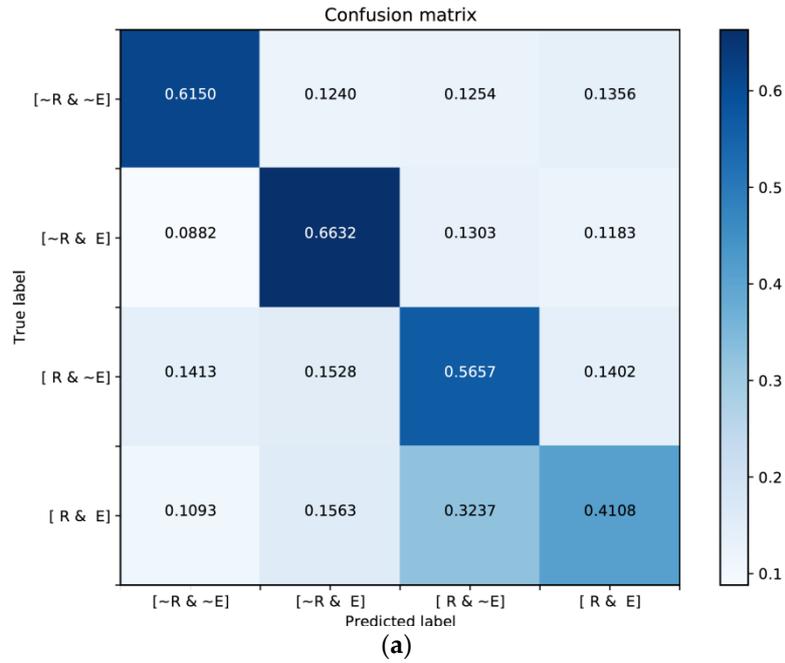

(a)

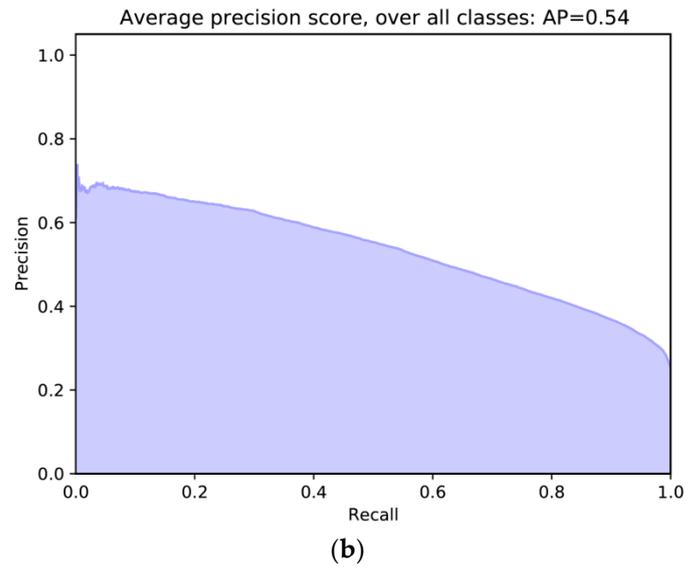

(b)

**Figure 9.** Model evaluation: (**a**) Normalized confusion matrix; (**b**) micro-averaged precision-recall curve for the four classes and its average precision (AP) score for the test data.

Table 7. Comparison of AP values between our model and random classifiers.

|  | AP of our classifier | AP of random classifier |
| --- | --- | --- |
| *not robust & not evolvable* | 0.52 | 0.135 |
| *not robust & evolvable* | 0.64 | 0.557 |
| *robust & not evolvable* | 0.81 | 0.238 |
| *robust & evolvable* | 0.17 | 0.071 |

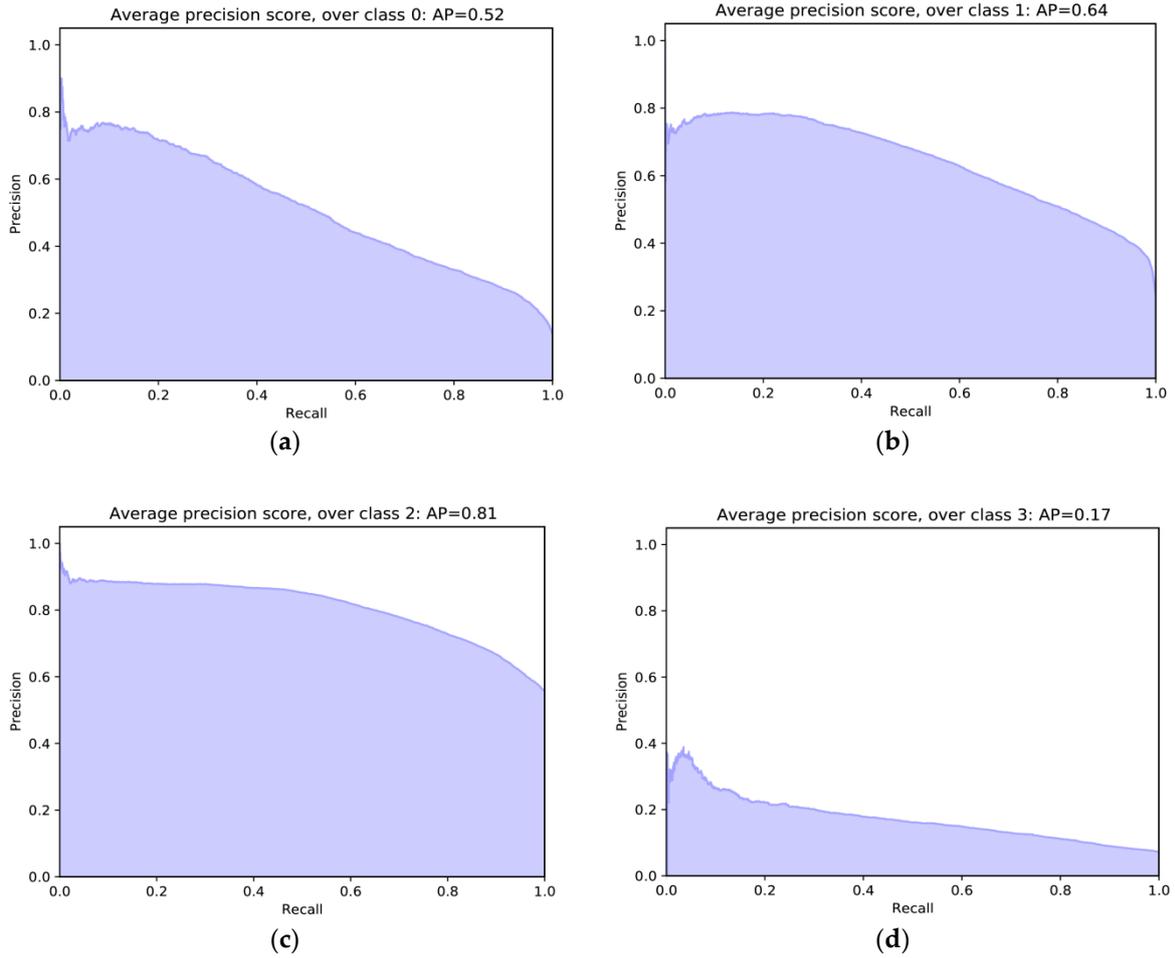

**Figure 10.** Precision-recall curves and average precision (AP) scores for the four classes for the test data: (**a**) *not robust & not evolvable*; (**b**) *not robust & evolvable*; (**c**) *robust & not evolvable*; (**d**) *robust & evolvable*.

## 4. Discussion

In this study, we classified robustness and evolvability in Boolean network models of biological systems into the four classes: *not robust & not evolvable, not robust & evolvable, robust & not evolvable*, and *robust & evolvable*. The classification was defined based on the change of attractors representing cell fates or cell functions before and after mutations. Using antifragility which is simply calculated from the dynamics during state transitions following external perturbations, we proposed a classifier to predict the properties of the four classes. We used a Convolutional Neural Network where the input is the difference of antifragility between original networks and their mutated ones and the output is the four classes. Our model showed a good performance for the multi-class classification. It indicates that our antifragility measure can play a role of a predictor to estimate the robustness and evolvability of biological networks.

To evaluate the utility of our antifragility in terms of a computational cost, we measured computation time of finding attractors and calculating antifragility. When searching for all attractors, we used Dubrova and Teslenko's algorithm which is a SAT-based approach to find attractors efficiently in synchronous Boolean networks [75]. This algorithm has two steps to find attractors. In the first step, a problem is encoded as input for the SAT solver. In the second step, the problem is solved by the SAT solver. However, SAT is NP-complete so the NP-complete problem is intractable in the worst case senario due to exponential complexity in the encoding or solving phase. For the 37 biological networks with less than 30 nodes in our study, finding attractors was faster than calculating antifragility (S3 in Supplementary Material). However, we tried to compute the attractors and antifragility of *lymphocytic leukemia* with 91 nodes ($2^{91}$ possible states) collected from *Cell Collective*. While it took about 47 minutes to calculate its antifragility value, we could not get its attractors even after a few days in our cluster equipped with 132G memory. Like *lymphocytic leukemia*, in some large Boolean networks, finding all attractors is computationally expensive or even unfeasible because of very long attractor lengths or a large number of attractors as the state space size is exponentially increased by the growing number of nodes. As the state space size is $2^N$, memory simply runs out if one attempts to explore exhaustively large networks. In this case, our classifier with antifragility can be a useful tool for studying the robustness and evolvability of biological networks, as it requires only a sample of the dynamics of original and perturbed networks.

For further study, we plan to use more data. We will collect more kinds of biological Boolean networks, introduce artificial data generated from random Boolean networks and add more mutations to the networks. Because the amount of data is one of important factors to have an influence on model performance, we will run simulations with more extensive data to find the better classifier. Besides, we will thoroughly explore the relationship between antifragility and robustness/evolvability. One possibility is that because antifragility reflects the characteristics of basins of attraction (*e.g.*, the structure of basins of attraction) and attractors (*e.g.*, the number and length of attractors), it might work for predicting the preservation and emergence of attractors. However, to reveal the relationship more clearly, we need further explorations. Taking a step forward, we will evolve random networks using antifragility as a fitness function and examine the robustness and evolvability of the resulting networks. This future research could give a clue to the mechanism of how biological systems obtained robustness and evolvability, and also suggest a possibility for developing robust and/or evolvable engineered systems which have antifragility as a control parameter.